\documentclass[revtex,prb,twocolumn,superscriptaddress,showpacs]{revtex4-1}

\usepackage{amsmath}   
\usepackage{amssymb}
\usepackage{dsfont}
\usepackage{braket}
\usepackage{natbib}
\usepackage{leftidx}
\usepackage{graphicx}    
\usepackage{verbatim}   
\usepackage{color}         
\usepackage{subfigure}  
\usepackage{hyperref}   
\usepackage{dcolumn}    
\usepackage{textcomp}
\usepackage{threeparttable}
\usepackage{titlecaps}
\usepackage{multirow}
\usepackage{lipsum}
\begin{document}

\title{Electronic structure of semiconductor nanostructures: A modified localization landscape theory}

\author{D. Chaudhuri}
\email{debapriya.chaudhuri@tyndall.ie} \affiliation{Tyndall National
Institute, University College Cork, T12 R5CP, Ireland}

\author{J.~C. Kelleher}
\affiliation{Department of Physics, University College Cork, Cork,
T12 YN60, Ireland}
\author{M.~R. O'Brien}
\affiliation{Tyndall National Institute, University College Cork,
T12 R5CP, Ireland} \affiliation{Department of Physics, University
College Cork, Cork, T12 YN60, Ireland}
\author{E~P. O'Reilly}
\affiliation{Tyndall National Institute, University College Cork,
T12 R5CP, Ireland} \affiliation{Department of Physics, University
College Cork, Cork, T12 YN60, Ireland}
\author{S. Schulz}
\affiliation{Tyndall National Institute, University College Cork,
T12 R5CP, Ireland}

\date{\today}

\begin{abstract}
In this paper we present a modified localization landscape theory to
calculate localized/confined electron and hole states and the
corresponding energy eigenvalues without solving a (large)
eigenvalue problem. We motivate and demonstrate the benefit of
solving $\hat{H}^2u=1$ in the modified localization landscape theory
in comparison to $\hat{H}u=1$, solved in the localization landscape
theory. We detail the advantages by fully analytic considerations
before targeting the numerical calculation of electron and hole
states and energies in III-N heterostructures. We further discuss
how the solution of $\hat{H}^2u=1$ is used to extract an effective
potential $W$ that is comparable to the effective potential obtained
from $\hat{H}u=1$, ensuring that it can for instance be used to
introduce quantum corrections to drift-diffusion transport
calculations. Overall, we show that the proposed modified
localization landscape theory keeps all the benefits of the recently
introduced localization landscape theory but further improves
factors such as convergence of the calculated energies and the
robustness of the method against the chosen integration region for
$u$ to obtain the corresponding energies. We find that this becomes
especially important for here studied $c$-plane InGaN/GaN quantum
wells with higher In contents. All these features make the proposed
approach very attractive for calculation of localized states in
highly disordered systems, where partitioning the systems into
different subregions can be difficult.

\end{abstract}


\maketitle

\section{Introduction}
Over the past two decades the calculation of the electronic
structure of semiconductor nanostructures such as quantum wells
(QWs) and quantum dots (QDs) has attracted enormous
attention.~\cite{DiCz98,StGr98,FrIt2000,BaGa2004,ChAn2005,WiSc2006,ScWi2007,MaMo2008,SiBe2009,ScCa2015}
This stems on the one hand from understanding and tailoring their
fundamental electronic and optical properties. On the other hand,
insight gained into the fundamental properties are also key for
optimizing or designing devices with new or improved characteristics
and capabilities. Energy efficient light emitting diodes (LEDs) are
amongst such devices.~\cite{AnRo2012,PiLi2017,Piprek2017} However,
from an atomistic standpoint, to model the single-particle states of
QDs, multi-QW (MQWs) or even full LED structures, the
(time-independent) Schr\"odinger equation (SE) has to be solved for
systems that can easily contain up to several million
atoms.~\cite{BeNa2003,ScCa2015} Given the large number of atoms,
standard density functional theory cannot be applied and empirical
models have been widely
used.~\cite{ScWi2007,MaMo2008,SiBe2009,ScCa2015} Even when employing
these more empirical models, in general, large eigenvalue problems
have to be solved, which can numerically still be demanding. The
numerical effort is even further amplified when calculations have to
be performed self-consistently, as for instance when describing
transport properties of LED structures.~\cite{LiPi2017}

Recently, and originally used to describe Anderson localization in
disordered systems, a new approach has been introduced in the
literature, which circumvents solving a large eigenvalue problem to
obtain (ground state) wave functions and energies of, for instance,
a QW. This approach is the so-called localization landscape theory
(LLT).~\cite{FiMa2012,ArDa2016,FiPi2017} Here, instead of solving
the time-independent SE, $\hat{H}\psi=E\psi$, and thus a (large)
eigenvalue problem, the idea is to solve
\begin{equation}
\hat{H}u=1\,\,. \label{eq:standardLL}
\end{equation}
The benefit of this approach is that only a set of linear equations
needs to be evaluated, which reduces the computational load
significantly, while giving results in very good agreement with the
solution of the time-independent SE.~\cite{FiPi2017} A detailed
analysis of the computational benefit of LLT can be found in
Ref.~\onlinecite{LiPi2017}, where ``standard'' self-consistent
SE-Poisson calculations for transport properties in InGaN/GaN-based
LEDs are compared to the results of a model that utilizes
drift-diffusion in combination with LLT. A speed up by a factor of
order 50 has been reported in Ref.~\onlinecite{LiPi2017} by the use
of the LLT based framework.

However, LLT is not only attractive from a numerical point of view,
it allows also to predict and capture physics that may be missing
in, for instance, semi-classical approaches. An example that was
mentioned already above and will be discussed in more detail below
is that it allows to establish quantum corrections to drift
diffusion models.~\cite{LiPi2017} Furthermore, LLT can be used to
describe Urbach tail energies observed in absorption spectra of
InGaN/GaN QW systems.~\cite{PiLi2017} Recently it has also been
applied to study localized vibrational modes in
enzymes.~\cite{ChPi2019} Finally, a recent development is also to
apply it to the Dirac equation for studying properties of graphene
or topological insulators.~\cite{LePa2019}

Taking all this together, LLT has several attractive advantages and
can give good agreement with the direct solution of the SE. However
care must be taken when calculating energies and eigenfunctions from
$u$. As described in Ref.~\onlinecite{FiPi2017}, the zero of energy
(reference energy) has to be carefully chosen to obtain good
agreement between energies $E$ calculated via LLT and SE.
Additionally, the region over which $u$ is integrated to obtain $E$
has to be selected carefully, as also demonstrated in
Ref.~\onlinecite{FiPi2017} for a single $c$-plane GaN/AlGaN QW. From
this, complications may arise in highly disordered systems, such as
InGaN wells with local variations in In content, where the system
has to be partitioned into ``appropriate'' regions to obtain
energies and wave functions that match closely the results obtained
by solving the SE.

Keeping all this in mind, here we describe a modified LLT (MLLT),
which keeps the benefits of the LLT, but has several advantages as
we will discuss and demonstrate below. Our starting point for the
MLLT is:
\begin{equation}
\hat{H}^2u=1\, . \label{eq:MLLT}
\end{equation}
Obviously the MLLT keeps the advantage of the LLT that instead of
solving a (large) eigenvalue problem, one is left with evaluating a
system of linear equations. Additionally, we will show that when
compared to LLT, the MLLT provides in general a better
description/faster convergence of the ground state energy with
respect to the SE results. This is especially true for higher In
contents. Also, we will demonstrate, by solving the SE, LLT and MLLT
numerically for electron and hole ground state energies in $c$-plane
InGaN/GaN QWs that the results of the MLLT are less sensitive to the
choice of the region over which $u$ is integrated to obtain these
energies.  Finally, we will discuss how to extract an effective
potential $W$ from MLLT that reflects and possesses similar features
as the effective potential obtained from LLT. This is important,
given that $W$ is for instance used in drift-diffusion studies of
InGaN/GaN QW-based LEDs, to account for quantum corrections in the
transport calculation frame.~\cite{LiPi2017} All this makes the MLLT
approach very attractive for studying for instance Anderson
localization or carrier transport in III-N based LEDs where
partitioning of the potential landscape in these highly disordered
systems might be difficult. We note that the MLLT approach was
discussed briefly in Ref.~\onlinecite{FiMa2012} but no detailed
study has yet been presented comparing the two approaches.

The manuscript is organized as follows. In Sec.~\ref{sec:Theory} we
briefly summarize aspects of the theoretical background of the LLT
which helps us to motivate the idea underlying the MLLT. In
Sec.~\ref{sec:PIAB} we apply LLT and MLLT to a particle-in-a-box
problem, since this allows us to flesh out fundamental aspects of
the LLT and MLLT approach. To further investigate fundamental
aspects and differences of LLT and MLLT, in an Appendix we briefly
investigate the solution of an infinite triangular well. This
analysis reveals that LLT diverges for this problem while MLLT
converges, but to a ground state energy that is noticeably different
from the SE solution. To apply LLT, MLLT along with the SE to
systems with a triangular but finite potential profile, we study
$c$-plane In$_{x}$Ga$_{1-x}$N/GaN single QWs. To do so, we first
introduce basic properties of III-N heterostructures in
Sec.~\ref{sec:BackgroundIIIN}. In Sec.~\ref{sec:Results} the results
from LLT and MLLT for $c$-plane In$_{x}$Ga$_{1-x}$N/GaN single QWs
are presented and compared to the solutions from the SE. We conclude
and summarize our work in Sec.~\ref{sec:Conclusion}.

\section{Localization landscape theory: Theoretical Background}
\label{sec:Theory}

In this section we present the theoretical background of our
studies. As already discussed in the introduction, the ``standard''
approach to calculate the electronic states and energies of
semiconductor heterostructures is based on solving the
time-independent SE:
\begin{equation}
\hat{H}{\psi_i} = E_i\psi_i\,\, . \label{eq:SE}
\end{equation}
Here, $\hat{H}$ is the Hamilton operator, $\psi_i$ the wave function
of state $i$ and $E_i$ the corresponding energy eigenvalue. To
calculate $E_i$ and $\psi_i$ for a system described by $\hat{H}$,
Eq.~(\ref{eq:SE}) can be treated as an eigenvalue problem. To do so,
the Hamiltonian matrix, corresponding to the Hamilton operator
$\hat{H}$ in Eq.~(\ref{eq:SE}), has to be constructed. The exact
form of this matrix depends on the choice of the underlying
electronic structure theory,~\cite{ScORE_book_2017} which for
semiconductor heterostructures usually ranges from empirical
pseudo-potential methods (EPM),~\cite{SiBe2009} to empirical
tight-binding models (ETBM),~\cite{CaSc2013local} over to
$\mathbf{k}\cdot\mathbf{p}$~\cite{StGr98} or single-band effective
mass approximations (EMA)~\cite{WoHa96}. The dimension of the
Hamiltonian matrix depends on several factors; in an atomistic
framework, such as EPM or ETBM, for instance on the number of atoms
in the system. Taking a Stranski-Krastanov grown QD as an example,
where both the dot and also the barrier material region have to be
taken into account in the theoretical modeling of its electronic
structure, several millions of atoms have to be
considered.~\cite{Best2009} As a consequence, one is left with a
large scale eigenvalue problem. Even though efficient numerical
routines are available, calculating the eigenstates and energies is
still demanding. The numerical burden further increases if
self-consistent calculations for optical properties, such as
self-consistent Hartree or Hartree-Fock calculations, are
required.~\cite{KiKa2010,PaSc2017}

To circumvent solving large eigenvalue problems, but at the same
time to gain insight into  wave functions and corresponding energies
of a quantum system, the LLT was introduced in 2012, especially
focusing on Anderson localization in highly disordered
systems.~\cite{FiMa2012} Recently this approach gained strong
interest for calculating the electronic structure of nitride-based
QW systems.~\cite{FiPi2017,PiLi2017,LiPi2017} Instead of evaluating
the SE, Eq.~(\ref{eq:SE}), LLT targets solving
Eq.~(\ref{eq:standardLL}), where $\hat{H}$ is again the Hamiltonian
operator of the system under consideration. As shown by Filoche
\emph{et al.},~\cite{FiPi2017} and as we will briefly outline below,
the function $u$ can be used to calculate the ground state energy
and wave functions of the system described by $\hat{H}$. This
summary of the LLT allows us also to motivate the MLLT.

As discussed in Ref.~\onlinecite{FiPi2017}, the function/state $u$
can be expressed in the basis formed by the eigenfunctions $\psi_i$
of $\hat{H}$:
\begin{equation}
|u\rangle=\sum_i\alpha_i|\psi_i\rangle\,\,  \label{eq:u_def}
\end{equation}
with
\begin{equation}
\alpha_i=\langle u|\psi_i\rangle=\iiint
u(\mathbf{r})\psi_i(\mathbf{r})\,d^3r\,\, .
\end{equation}
Due to the self-adjointness of $\hat{H}$, $\alpha_i$ can be obtained
via
\begin{equation}
\alpha_i=\langle u|\psi_i\rangle=\frac{1}{E_i}\langle
u|\hat{H}\psi_i\rangle=\frac{1}{E_i}\langle 1|\psi_i\rangle\,\, .
\label{eq:alpha}
\end{equation}
From Eq.~(\ref{eq:alpha}) one can see that contributions from
energetically higher lying states to $u$, Eq.~(\ref{eq:u_def}),
depend on the factor $1/E_i$. Therefore, if the energy separation
between state $i$ and $i+1$ is small, for example between the ground
state ($i=1$) and the first excited state ($i=2$), several states
may contribute significantly to the expansion in
Eq.~(\ref{eq:u_def}). This is obviously undesirable when $u$ should
approximate for instance the ground state wave function $\psi_1$
obtained from the SE.

Furthermore, assuming as an example a QW system, so that electron
and hole wave functions are localized in a subregion of the full
well-barrier system, the energetically lowest states contributing to
$u$ in Eq.~(\ref{eq:u_def}) are basically the fundamental, local
quantum states in this subregion. In many cases, for example when
looking at radiative recombination of carriers, these are the states
one is interested in. Therefore, in each localization subregion
$\Omega_m$, $u$ can be estimated from~\cite{FiPi2017}
\begin{equation}
|u\rangle\simeq\frac{\langle1|\psi^{m}_1\rangle}{E^{m}_1}|\psi^{m}_1\rangle=\alpha_1|\psi^{m}_1\rangle\,\,
,
\end{equation}
where $|\psi^{m}_1\rangle$ is the local fundamental state in the
subregion $\Omega_m$. Following Ref.~\onlinecite{FiPi2017},
$|\psi^{m}_1\rangle$ can be assumed to be proportional to $u$ in
subregion $\Omega_m$:
\begin{equation}
|\psi^{m}_1\rangle\approx\frac{|u\rangle}{||u||}\,\,.
\label{eq:psi_approx}
\end{equation}

Finally, using Eq.~(\ref{eq:psi_approx}) one can approximate the
fundamental/ground state energy in subregion $\Omega_m$ from:
\begin{eqnarray}
\nonumber E^{m}_1 &=&
\langle\psi^{m}_1|\hat{H}|\psi^{m}_1\rangle\approx \frac{\langle
u|\hat{H}|u\rangle}{||u||^2}=\frac{\langle
u|1\rangle}{||u||^2}\\
&=&\frac{\iiint_{\Omega_{m}}
u(\bold{r})d^{3}\bold{r}}{\iiint_{\Omega_{m}}
u^2(\bold{r})d^{2}\bold{r}}\, . \label{eq:Energy_LLT}
\end{eqnarray}
Thus, from this equation it is clear that the function
$u(\bold{r})=\langle r|u\rangle$ provides a direct estimate of the
(ground state) energy.

However, $u(\bold{r})$ is not only connected to the
ground state energy and wave function, it also defines an
``effective confining potential'', which is given by
$W=1/u$.~\cite{FiMa2012,FiPi2017} One can show that $W$ is related
to the exponential decay of localized states away from their (main)
localization subregion. This decay of the wave function is then
connected to tunneling effects, as shown for instance by the
Wenzel-Kramers-Brillouin (WKB) approximation. Therefore, the
effective confining potential $W$ has attracted interest for
drift-diffusion calculations, given that $W$ then introduces quantum
corrections into these semi-classical transport
models.~\cite{LiPi2017}

Taking all this together, several points can be concluded from the
above. First, to obtain $u$ and for instance $E^{m}_1$, the system
has to be partitioned into subregions $\Omega_m$ so that
Eq.~(\ref{eq:psi_approx}) is a good approximation. Secondly, the
reference energy or zero of energy should be chosen so that the
expansion of $|u\rangle$, Eqs.~(\ref{eq:u_def})
and~(\ref{eq:alpha}), respectively, is dominated by the expansion
coefficient $\alpha_1$. In other words contributions from
energetically higher lying states to $u$ are then of secondary
importance.

The last aspect motivates the modified LLT (MLLT),
Eq.~(\ref{eq:MLLT}), and is triggered by two factors. First, when
calculating eigenvalues and eigenfunctions of for instance
semiconductor QDs, very often the so-called folded spectrum method
(FSM) is applied to turn an interior eigenvalue problem into finding
the lowest energy eigenvalue.~\cite{WaZu94} More precisely, in the
FSM, instead of solving the eigenvalue problem $\hat{H}\psi=E\psi$,
one evaluates $(\hat{H}-\epsilon\mathds{1})^2\psi=\tilde{E}\psi$.
Here, $\mathds{1}$ is the unit operator and $\epsilon$ is the
so-called reference energy around which the spectrum is folded. In
case of $\epsilon=0$, $\tilde{E}=E^2$. Working with $\hat{H}^2$ for
the LLT, thus resulting in MLLT, has now the following advantages
for the expansion of $u$ in terms of $|\psi_i\rangle$. Using
Eq.~(\ref{eq:u_def}) the expansion coefficients $\alpha_i$ are given
by:
\begin{equation}
\alpha^\text{MLLT}_i=\langle u|\psi_i\rangle=\frac{\langle
u|\hat{H}^2\psi_i\rangle}{E^2_i}=\frac{\langle\hat{H}^2
u|\psi_i\rangle}{E^2_i}=\frac{\langle 1|\psi_i\rangle}{E^2_i}\,\, .
\label{eq:alpha_MLLT}
\end{equation}
Therefore:
\begin{equation}
|u\rangle=\sum_i\frac{\langle
1|\psi_i\rangle}{E^2_i}|\psi_i\rangle\,\, .
\label{eq:MLLT_u_expansion}
\end{equation}
As one can see from this equation, the contributions from higher
lying energy states come in with $1/E^2_i$ instead of $1/E_i$ as in
the ``standard'' LLT. Therefore, the here proposed MLLT should lead
to an even better approximation of the fundamental wave function in
a subregion and therefore a better approximation of the
corresponding energy.

While this clearly shows the benefit of using MLLT in calculating
wave functions and energies, the question is how to obtain the
effective potential $W$ from MLLT? Here, care must be taken since
$u$ itself has now the dimension inverse energy squared. To obtain
$W_\text{MLLT}$ from MLLT one can define $W_\text{MLLT}=(E_l\cdot
u_\text{MLLT})^{-1}$, where $E_l$ is for example the ground state
energy of the systems under consideration. However, as we see from
Eq.~(\ref{eq:MLLT_u_expansion}), several different energies may
contribute to the expansion of $u$. Another option is for instance
to define the effective potential $W_\text{MLLT}$ via
\mbox{$W_\text{MLLT}=(\sqrt{u_\text{MLLT}})^{-1}$}. Given the
importance of the effective potential $W$ for describing localized
states and also tunneling effects, it is therefore important to
analyze the effective potential in more detail and compare it to
$W_\text{LLT}$ obtained from``standard'' LLT.

To highlight and demonstrate the benefits of the
MLLT further for wave functions and energies, but also to gain
insight into $W_\text{MLLT}$, we first study a simple
particle-in-a-box problem with infinitely high barriers in the next
section. This calculation can be done fully analytically and offers
therefore a very transparent test case for the two methods and to
compare the results directly with the results from solving the SE.

\section{Localization landscape theory and modified localization
landscape theory: Application to a square well with infinitely high
barriers} \label{sec:PIAB}

In this section, we apply both LLT and MLLT to the simple
particle-in-a-box problem with infinitely high barriers, since here
fully analytic solutions can be derived. The benefit of this is
twofold: (i) it sheds light onto general features of the LLT and
(ii) it demonstrates the advantages of the proposed MLLT. We compare
the results obtained from LLT and MLLT with those from the SE.

We start with the SE and its solution for this problem. Assuming the
well boundaries to be at $z=0$ and $z=L$, and choosing the potential
energy to be zero for \mbox{$0<z<L$}, the SE in this region reads:
\begin{equation}
 -\frac{\hbar^2}{2m}\frac{d^2}{dz^{2}}\psi_{n}(z) =
 E_{n}\psi_{n}(z)\,\, .
\label{5}
\end{equation}
For $z\leq0$ and $z\geq L$ the potential energy is infinitely large.
Due to the boundary conditions $\psi_n(0) = 0$ and $\psi_n(L) = 0$,
the eigenvalues $E_n$ and the normalized eigenstates $\psi_n(z)$ are
given by:~\cite{Nolting_2017}
\begin{equation}
E_n=\frac{n^2\pi^2\hbar^2}{2mL^2}
\label{7}
\end{equation}
and
\begin{equation}
\psi_{n}(z) = \sqrt{\frac{2}{L}}\sin\left(\frac{n\pi
z}{L}\right)\,\, . \label{eq:EF_Infinite_SE}
\end{equation}
The ground state energy eigenvalue $E_1=\pi^2\hbar^2/2mL^2$ will now
serve as a reference for our calculations using LLT and MLLT,
respectively.

\subsection{LLT solution}

Following Eq.~(\ref{eq:u_def}), $u(z)$ can be expressed as a linear
combination of the eigenfunctions $\psi_n(z)$,
Eq.~(\ref{eq:EF_Infinite_SE}), which form a complete basis set for
the Hilbert space:
\begin{equation}
u = \sum_{n}\alpha_n\psi_n\,\, . \label{9}
\end{equation}
Now, exploiting the LLT equation $\hat{H}u=1$ and using
Eq.~(\ref{eq:EF_Infinite_SE}), from Eq.~(\ref{9}) we obtain:
\begin{eqnarray}
\hat{H}u = \sum_{n}E_{n}\alpha_n\psi_n = 1 \nonumber \\
\Rightarrow\sum_{n}\alpha_n E_1
n^2\sqrt{\frac{2}{L}}\sin\left(\frac{n\pi z}{L}\right) =1\,\, .
\label{eq:LLT_psi}
\end{eqnarray}
Here, we recall that the constant function 1 can be represented
by~\cite{Fogiel1991}
\begin{equation}
\frac{4}{\pi}\sum_{n_\text{odd}}^{\infty}\frac{1}{n}\sin\left(\frac{n\pi
z}{L}\right) =1\,\, . \label{eq:constant_1}
\end{equation}
Thus, combining Eq.~(\ref{eq:LLT_psi}) and~(\ref{eq:constant_1}),
the coefficients $\alpha_n$ are zero for even $n$ and for the odd
values of $n$ they read:
\begin{equation}
\alpha_n = \frac{2\sqrt{2L}}{E_1\pi n^3}\,\, . \label{13}
\end{equation}
Using this expression for $\alpha_n$, Eq.~(\ref{13}), $u$,
Eq.~(\ref{9}), is therefore given by:
\begin{align}
u(z) &= \sum_{n_{odd}}^{\infty}\frac{2\sqrt{2L}}{E_1n^3\pi}\psi_n(z) \nonumber \\
       &= \frac{2\sqrt{2L}}{E_1\pi}\sum^{\infty}_{m=1}\frac{\psi_{2m-1}(z)}{(2m-1)^3} \nonumber \\
       &= \lambda\left(\psi_{1}(z) + \frac{1}{27}\psi_3(z) + \frac{1}{125}\psi_5(z) +
       ...\right)\,\, ,
\label{eq:u_expansion_PIB}
\end{align}
where $\lambda=\frac{2\sqrt{2L}}{E_1\pi}$. From this equation it
follows that the series expansion of $u$ converges as $1/n^3$ with
significantly lesser contributions from the higher order terms.
Furthermore, only every second basis state of the infinite square
well eigenstates contributes. Thus, for this problem, the LLT gives
a very good approximation of the ground state wave function
$\psi_1(z)$, but, since for instance $\psi_{2}(z)$ is missing in the
expansion, the first excited state cannot be described by $u(z)$ in
general. However, we highlight here that when applying LLT to
disordered systems where several minima/subregions $\Omega_m$ can be
defined, LLT can be applied to the different $\Omega_m$ and one can
find the fundamental state for each subregion. While locally this is
the ground state, globally these states will be excited
states.~\cite{ArDa2019} In addition, an analysis based on Weyl's Law
has shown that the LLT can give a very good estimate of the
integrated density of states over a significant energy range,
despite that it cannot be used to estimate individual higher state
energies in a given local minimum.~\cite{ArDa2019} Turning back to
our problem here, $u$ gives a very good description of the
fundamental state in the subregion $\Omega_m=[0,L]$. It is important
to remember that the $1/n^3$ convergence resulted directly from
$\hat{H}u=1$. So the MLLT approach, utilizing $\hat{H}^2u=1$ should
lead to an even faster convergence of the series expansion of $u$ in
terms of the eigenstates $\psi_n(z)$ in the subregion
$\Omega_m=[0,L]$. Before discussing this in more detail we turn and
calculate the ground state energy of the one-dimensional (1-D)
infinite square well potential problem within LLT.

\begin{figure}[t]
\includegraphics[width=8.6cm, clip=true]{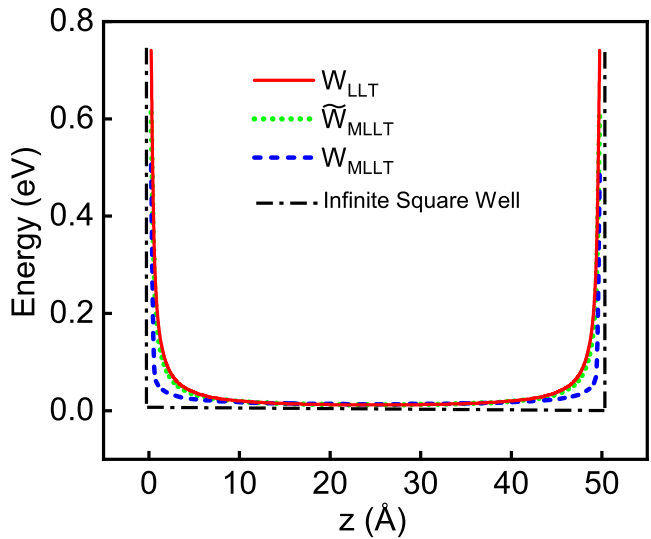}
\caption{(Color online) Comparison of the effective potentials for a
square well with infinitely high potential barriers. The infinite
square well potential is given by the (black) dashed dotted line.
The effective confining potential calculated via LLT is given by the
red solid line. Effective confining potentials obtained from MLLT
via two different approaches (see main text) are given by the (blue)
dashed and (green) dotted line.} \label{1a}
\end{figure}

Using Eq.~(\ref{eq:Energy_LLT}), and keeping in mind
$\langle\psi_n|\psi_m\rangle=\delta_{n,m}$, the energy
$E^{m}_{1,\text{LLT}}$ is given by:
\begin{align}
E^{m}_{1,\text{LLT}} &=  \frac{\bra{u}H\ket{u}}{||u||^2}= \frac{\langle u\ket{1}}{||u||^2}\nonumber \\
       &=
       \frac{2\lambda\sqrt{2L}}{\pi\lambda^2}\frac{{\sum_{m=1}^\infty\frac{1}{(2m-1)^4}}}{\sum_{m=1}^\infty\frac{1}{(2m-1)^6}}=\frac{2\sqrt{2L}}{\lambda\pi}\frac{10}{\pi^2}\,\,
       .
\label{15}
\end{align}
Substituting the value of $\lambda=\frac{2\sqrt{2L}}{E_1\pi}$ into
Eq.~(\ref{15}) one is left with
\begin{align}
E^{m}_\text{1,LLT} \approx 1.0132\cdot E_1\,\, . \label{16}
\end{align}
Thus, the ground state or fundamental energy $E^{m}_\text{1,LLT}$ in
the subregion $\Omega_m=[0,L]$ is in excellent agreement with the
result obtained directly from the SE; $E^{m}_\text{1,LLT}$ is just
over 1\% larger than the ground state energy eigenvalue $E_1$.
However, as we will discuss below and in an appendix, it is not
guaranteed that always such a good agreement is achieved and that
LLT might even fail for certain confinement potentials.

Having discussed energy eigenvalues, we turn now to consider the
effective potential $W_\text{LLT}$ resulting from the LLT. This is
given by $W_\text{LLT}=u^{-1}$ and shown in Fig.~\ref{1a} by the red
solid line along with the potential of a square well (black dashed
dotted line) of width 50 \AA{} and with infinitely high barriers.
Figure~\ref{1a} shows that $W_\text{LLT}$ softens the potential near
the boundaries. As we will discuss further below, this effect is
also seen in a well with finite barriers, where it provides the
above discussed quantum corrections to transport
simulation.~\cite{LiPi2017} Therefore, it is important that the MLLT
captures these pertinent aspects as well, to be of use for such
simulations. In the following section we discuss the MLLT for a
square well with infinitely high barriers.

\subsection{MLLT}

Having solved the particle-in-a-box problem within the SE and LLT, we
target it now within MLLT by employing:
\begin{equation}
\hat{H}^2u=1\,\,. \label{eq:MLLT_2}
\end{equation}
Using Eq.~(\ref{eq:u_def}), in the case of the MLLT one is left with
\begin{equation}
\hat{H}^2u =\sum_{n}E_n^2\alpha_n\psi_n = 1\,\, . \nonumber \\
\label{17}
\end{equation}
Following the steps outlined above for the LLT, the expansions
coefficients $\alpha_n$, again taking only odd $n$ values, are given
by
\begin{equation}
\alpha_n = \frac{2\sqrt{2L}}{E_1^2\pi n^5}\,\, . \label{20}
\end{equation}
Comparing the above equation with Eq.~(\ref{13}), we find here already
that $\alpha_n$ scales as $1/n^5$ instead of $1/n^3$. With this $u$
reads:
\begin{align}
u(z) &= \frac{2\sqrt{2L}}{E_1^2\pi}\sum_{n_\text{odd}}^{\infty}\frac{1}{n^5}\psi_n(z) \nonumber \\
       &= \lambda'\sum^{\infty}_{m=1}\frac{\psi_{2m-1}(z)}{(2m-1)^5} \nonumber \\
       &= \lambda'\left(\psi_{1}(z) + \frac{1}{243}\psi_3(z) + \frac{1}{3125}\psi_5(z) +
       ...\right)\,\, ,
\label{21}
\end{align}
with $\lambda'=2\sqrt{2L}/E_1^2\pi$. When comparing this result with
the expansion of $u$ in the LLT frame,
Eq.~(\ref{eq:u_expansion_PIB}), it is evident that the MLLT yields
an even faster/better convergence/approximation of $u(z)$ with
respect to the ground state/fundamental state $\psi_1$. Thus, within
the MLLT approach the approximation $\psi^{m}_1\approx u/||u||$,
Eq.~(\ref{eq:psi_approx}), should be even better justified.

The \emph{square} of the energy eigenvalue $(E^m_\text{1,MLLT})^2$
is given by:
\begin{align}
(E^m_\text{1,MLLT})^{2} &=  \frac{\bra{u}H^2\ket{u}}{||u||^2}= \nonumber \\
       &= \frac{2\sqrt{2L}}{\pi\lambda'}\frac{{\sum_{m=1}^\infty\frac{1}{(2m-1)^6}}}{\sum_{m=1}^\infty\frac{1}{(2m-1)^{10}}}\\
       &\approx 1.001\cdot E_1^2\,\, .  \nonumber 
\end{align}
Therefore, $E^m_\text{1,MLLT}\approx1.0005\cdot E_1$ yields an even
better approximation of the true ground state energy, when compared
to the LLT result discussed above
\mbox{($E^m_\text{1,LLT}\approx1.0132\cdot E_1$)}. Again, the reason
for this improvement can be traced back to the series expansion of
$u$ where the expansion coefficients $\alpha_n$ decrease rapidly in
magnitude with increasing $n$.

Having seen the improved ground state energy convergence in MLLT, we
now turn our attention to the calculation of the effective confining
potential $W_\text{MLLT}$ within MLLT. In the previous section we
have already discussed two approaches to obtain $W_\text{MLLT}$ from
$u_\text{MLLT}$, namely \mbox{$\widetilde{W}_\text{MLLT}=(E_l\cdot
u_\text{MLLT})^{-1}$} or
\mbox{$W_\text{MLLT}=(\sqrt{u_\text{MLLT}})^{-1}$}. From
Eq.~(\ref{21}), it is clear that
\mbox{$\widetilde{W}_\text{MLLT}=(E_l\cdot u)^{-1}$} with $E_l=E_1$
will give an effective potential $\widetilde{W}_\text{MLLT}$ that
will be in excellent agreement with $W_\text{LLT}$, given that
$\lambda'=2\sqrt{2L}/(E^2_1\pi)$. This is confirmed in
Fig.~\ref{1a}, where $\widetilde{W}_\text{MLLT}$ (green dashed line)
matches almost perfectly $W_\text{LLT}$, thus keeping the feature of
softening the potential at the infinitely high barriers. Also the
second approach,
$\widetilde{W}_\text{MLLT}=(\sqrt{u_\text{MLLT}})^{-1}$ gives a
reasonable description of the potential, however, with a less
pronounced softening near the barrier.

Overall, we see for the infinite square well potential that
$\widetilde{W}_\text{MLLT}=(E_1\cdot u)^{-1}$ gives an effective
potential that matches closely $W_\text{LLT}$, reflecting that each
expansion coefficient $\alpha_n$, cf. Eq.~(\ref{20}), only depends
on the ground state energy $E_1$. However, as indicated already in
Sec.~\ref{sec:Theory}, Eq.~(\ref{eq:alpha_MLLT}), this might not be
the case for other potentials. We discuss this further briefly in
the appendix, where we apply the MLLT to a triangular shaped well
with \emph{infinitely} high barriers. For such a potential we find
that LLT does not converge to give a finite estimate of the ground
state energy $E_1$; MLLT does converge but to an energy that is
noticeably different from the solution of the SE. Given that both
LLT and MLLT have difficulties in dealing with a triangular shaped
potential with infinite barriers, we investigate a triangular shaped
well with \emph{finite} barriers in the following. Such a system is
relevant for studying electronic and optical properties of
III-N-based QW systems, as we describe in the next section.

\begin{table}[t]
\begin{ruledtabular}
\centering \caption{Band gap $E_g$,~\cite{CaSc2011} lattice
constants $a$,$c$,~\cite{CaSc2011}spontaneous polarization
$P_\text{sp}$,~\cite{CaSc2011} piezoelectric coefficients
$e_{ij}$,~\cite{CaSc2011} elastic constants $C_{ij}$~\cite{CaSc2011}
and effective electron $m_e$~\cite{RiWi2008} and hole $m_{h}$ mass
for wurtzite InN and GaN. The hole mass has been determined from the
equations given in Ref.~\onlinecite{ScBa2010}, using the
$A_i$-parameters from Ref.~\onlinecite{RiWi2008}.}
\label{tab:Matpara}
\begin{tabular}{l c c c  c }
Parameters & & GaN & &InN  \\ [0.5ex]
\hline
$E_g$ (eV)  & &3.44 & &0.64  \\
$a$ (\AA) & &3.189 &&3.545 \\
$c$ (\AA)& &5.185 & &5.703 \\
$P_\text{sp}$ (C/$m^2$)& &-0.034 & & -0.042 \\
$e_{31}$ (C/$m^2$)& &-0.45& & -0.52 \\
$e_{33}$ (C/$m^2$)& &0.83& & 0.92 \\
$C_{13}$ (GPa)& &106& & 92 \\
$C_{33}$ (GPa)& &398 & & 224 \\
$m_{e}(m_0)$   & &0.209 & & 0.068     \\
$m_{h}(m_0)$   & &1.876 & & 1.811     \\
\end{tabular}
\end{ruledtabular}
\end{table}

\section{Background on nitride-based heterostructures and $\text{In}\text{Ga}$N Quantum Well}
\label{sec:BackgroundIIIN}

III-N materials, such as InN, GaN and AlN have attracted
considerable interest for optoelectronic devices, since their alloys
are in principle able to cover emission wavelengths from infrared to
deep ultra-violet.~\cite{Hump2008} InGaN heterostructures, such as
QWs, are of particular interest for emission in the visible spectral
range.~\cite{Hump2008} When compared to other III-V materials, such
as InAs or GaAs, III-N materials preferentially crystallize in the
wurtzite crystal phase while InAs and GaAs crystallize in the zinc
blende phase. The wurtzite crystal structure, due to its lack of
inversion symmetry, allows for a strain-induced piezoelectric
polarization vector field but also a spontaneous polarization vector
field, which is even present in the absence of any strain
effects.~\cite{BeFi97} Discontinuities in the polarization vector
fields lead to very strong electrostatic built-in fields (MV/cm) in
InGaN/GaN QW systems grown along the wurtzite $c$-axis, which is the
standard growth direction for these
systems.~\cite{ScKn2007,PaSc2017} Often these systems, especially
when dealing with transport properties of InGaN/GaN MQW-based LED
structures, are treated as 1-D systems in which the conduction and
valence band profiles are modified by the presence of the intrinsic
electrostatic built-in potentials.~\cite{ScKn2007,CaSc2011} It
should be noted that this is a simplified description of these
systems; more recently it has been shown that the alloy
microstructure of InGaN QWs significantly affects the electronic
structure, so that local potential fluctuations play an important
role.~\cite{WaGo2011,ScCa2015} However, for the analysis here, a
simplified 1-D model is a good starting point for comparing LLT and
MLLT and to highlight the benefits of the MLLT in terms of
convergence and ``robustness'' of the solution against, for
instance, the choice of the sub-region $\Omega_m$ over which $u$ is
being evaluated to obtain the ground state energy in the given
region. Since the methodology of the MLLT is the same as that of the
LLT, MLLT can directly be applied to a landscape with energy
fluctuations due to alloy fluctuations. However, as discussed
already above, further consideration must then be given as to how
best to calculate $W$ within MLLT, which we will do below. To flesh
out the benefits of the MLLT, we focus on the often used 1-D
description of the electronic structure of $c$-plane
In$_x$Ga$_{1-x}$N/GaN QWs with different In contents $x$.
As mentioned above, due to the underlying wurtzite crystal structure
and growth along the $c$-axis, $c$-plane InGaN/GaN QW systems
exhibit very strong electrostatic built-in fields. This
electrostatic field arises from discontinuities in spontaneous and
piezoelectric polarization vector fields. The corresponding total
built-in potential $\phi^\textnormal{{QW}}$, assuming that the
wurtzite $c$-axis is parallel to the $z$-axis of the coordinate
system, can be expressed as:~\cite{CaSc2011}
\begin{align}
\label{eq:phi_piezo_spon}
\phi^\textnormal{{QW}}(z) &= \phi^\textnormal{{QW}}_{\textnormal{sp}}(z) + \phi^\textnormal{{QW}}_{\textnormal{pz}}(z) \nonumber \\
                     &= \Bigg\{\frac{(P_{sp}^W - P_{sp}^B)+P_{pz}^W}{2\epsilon_0\epsilon^W_r}
                     \Bigg\}(|z|-|z-h|)\,\, .
\end{align}
Here, $h$ is the height/width of the QW with well barrier interfaces
at $z$=0 and $z$ = $h$. The dielectric constant of the QW material
is denoted by $\epsilon^W_r$ and $P_{sp}^W$ ($P_{sp}^B$) is the
spontaneous polarization in the well (barrier). Assuming that the
barrier material is strain-free, a strain field is only present in
the InGaN QW, since InGaN has a larger lattice constant than
GaN.~\cite{VuMe2003} Thus one is left with a piezoelectric
polarization component in the well $P_{pz}^W$, which in the 1-D case
can be written as:~\cite{ScCa2011}
\begin{align}
P_{pz}^W = 2 \epsilon_{11}e_{31}^W + \epsilon_{33}e_{33}^W\,\, .
\end{align}
Here, $e^{W}_{ij}$ and $\epsilon_{ij}$ are the (well) piezoelectric
coefficients and the strain tensor components. The strain tensor
components are given by
\mbox{$\epsilon_{33}=(-2{C_{13}^W}/{C_{33}^W})\epsilon_{11}$} and
\mbox{$\epsilon_{11}=(a^B-a^W)/a^W$}; $a^{W}$ ($a^{B}$) is the
in-plane lattice constant of the well (barrier) material and
$C^{W}_{ij}$ are the elastic constants of the well material. The
material parameters used in this study are summarized in
Tab.~\ref{tab:Matpara}. When calculating the electrostatic built-in
potential of In$_x$Ga$_{1-x}$N/GaN QWs as a function of the In
content $x$, a linear interpolation of the involved material
parameters is applied. We neglect contributions from second-order
piezoelectric effects.~\cite{PaSc2017} Using
Eq.~(\ref{eq:phi_piezo_spon}) and the material parameters from
Tab.~\ref{tab:Matpara}, the resulting built-in potential is similar
to that of a capacitor.~\cite{CaSc2011}

Since we are interested in a general comparison between LLT and MLLT
results, we calculate the electronic structure of the above
discussed In$_{x}$Ga$_{1-x}$N/GaN QW systems in the framework of a
single-band effective mass approximation for electrons and holes.
The confining potential for electrons and holes is then given by the
conduction band (CB) and valence band (VB) edge alignment between
GaN and InGaN. In Eq.~(\ref{eq:BOset}) below we assume that the VB
edge of bulk GaN (no built-in field) denotes the zero of energy in
our system; the GaN CB edge, in the absence of the built-in field,
is at the band gap energy $E^\text{GaN}_g$ of bulk GaN. The
In$_x$Ga$_{1-x}$N CB edge, $E^\text{InGaN}_{\textnormal{CB}}$, and
the VB edge, $E^\text{InGaN}_{\textnormal{VB}}$, are calculated as a
function of the In content $x$ as follows:~\cite{CaSc2013local}
\begin{eqnarray}
E^\text{InGaN}_{\textnormal{CB}} &=& x(E_g^{\textnormal{InN}} + \Delta{E_{\textnormal{VB}}}) + (1-x)E_g^{\textnormal{GaN}}\nonumber\\
& & - b_{\textnormal{CB}}x(1-x)\,\, , \nonumber \\
E^\text{InGaN}_{\textnormal{VB}} &=& x\Delta{E_{\textnormal{VB}}} -
b_{\textnormal{VB}}x(1-x)\,\, . \label{eq:BOset}
\end{eqnarray}
Here, $\Delta{E_{\textnormal{VB}}}$ is the natural VB offset between
pure InN and GaN, which has been taken from HSE-DFT
calculations.~\cite{MoMi2011} The (composition dependent) CB and VB
edge bowing parameters are denoted by $b_{\textnormal{CB}}$ and
$b_{\textnormal{VB}}$.~\cite{CaSc2013local} In combination with the
built-in potential from above, the band edge profile shows the well
known triangular-shaped profile, leading to the situation that
electrons and holes are spatially separated along the growth
direction ($c$-axis/$z$-axis). This situation is also know as the
quantum confined Stark effect (QCSE).~\cite{ImKo98}

Building on this potential profile we use a single-band effective
mass approximation to construct the Hamiltonian matrix of this
system. Here, we use different effective masses for electrons and
holes, with the values given in Table~\ref{tab:Matpara}. A linear
interpolation between the effective masses of InN and GaN has been
applied to obtain the corresponding masses for InGaN.  However,
differences in the effective mass inside and outside the well are
not considered. Given that we are interested in ground state
energies and more generally in a comparison between results obtained
from the SE, LLT and MLLT, applying a constant effective mass should
be sufficient for our purposes here. To numerically solve
$\hat{H}\psi=E\psi$, $\hat{H}u=1$ and $\hat{H}^2u=1$ we use the
finite difference method and assume a well of width of $L_w=35$
\AA\, with a barrier width of $L_{b}=100$ \AA\ on each side of the
well; the discretization step size is $\Delta=0.05$ \AA.

\section{Results for $c$-plane $\text{In}_x\text{Ga}_{1-x}$N/$\text{Ga}$N quantum wells}
\label{sec:Results}

In this section we compare and discuss ground state energies, wave
functions and the effective potential $W$ of $c$-plane
In$_x$Ga$_{1-x}$N/GaN QWs obtained by solving SE, LLT and MLLT in
the numerical framework discussed above. Special attention is paid
to the impact of the In content $x$ on the results, given that with
increasing In content the piezoelectric contribution to the built-in
potential, and thus the ``tilt'' in the band edge profiles,
increases. More specifically, we study here In contents $x$ ranging
from 5\% up to 50\%, even though the very high In contents
($x>25\%$) are experimentally very difficult to achieve for fully
strained $c$-plane wells. Such an analysis will help us to compare
the impact of strong asymmetries in the potential landscape on the
results of the LLT and MLLT, respectively. Strong fluctuations in
the potential landscape may occur locally in InGaN QWs with higher
In contents (e.g. 25\% In) due to random alloy fluctuations.

\begin{figure}[t!]
\includegraphics[width=8.6cm, clip=true]{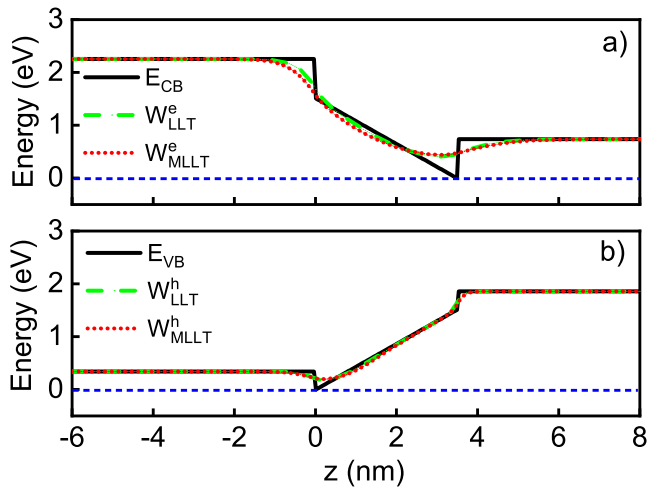}
\caption{(Color online) a) Conduction $E_\text{CB}$ (black solid
line) and b) valence band edge $E_\text{VB}$ (black solid line) in a
$c$-plane InGaN/GaN QW with 25 $\%$ In. The effective potentials
calculated from LLT, $W_\text{LLT}$, and MLLT, $W_\text{MLLT}$, are
given by the green dashed lines and the red dotted lines,
respectively. More details on the calculation of $W_\text{MLLT}$ are
given in the main text.} \label{fig:Band_edge}
\end{figure}
\begin{figure*}[t!]
\centering
\includegraphics[width=\textwidth, clip=true]{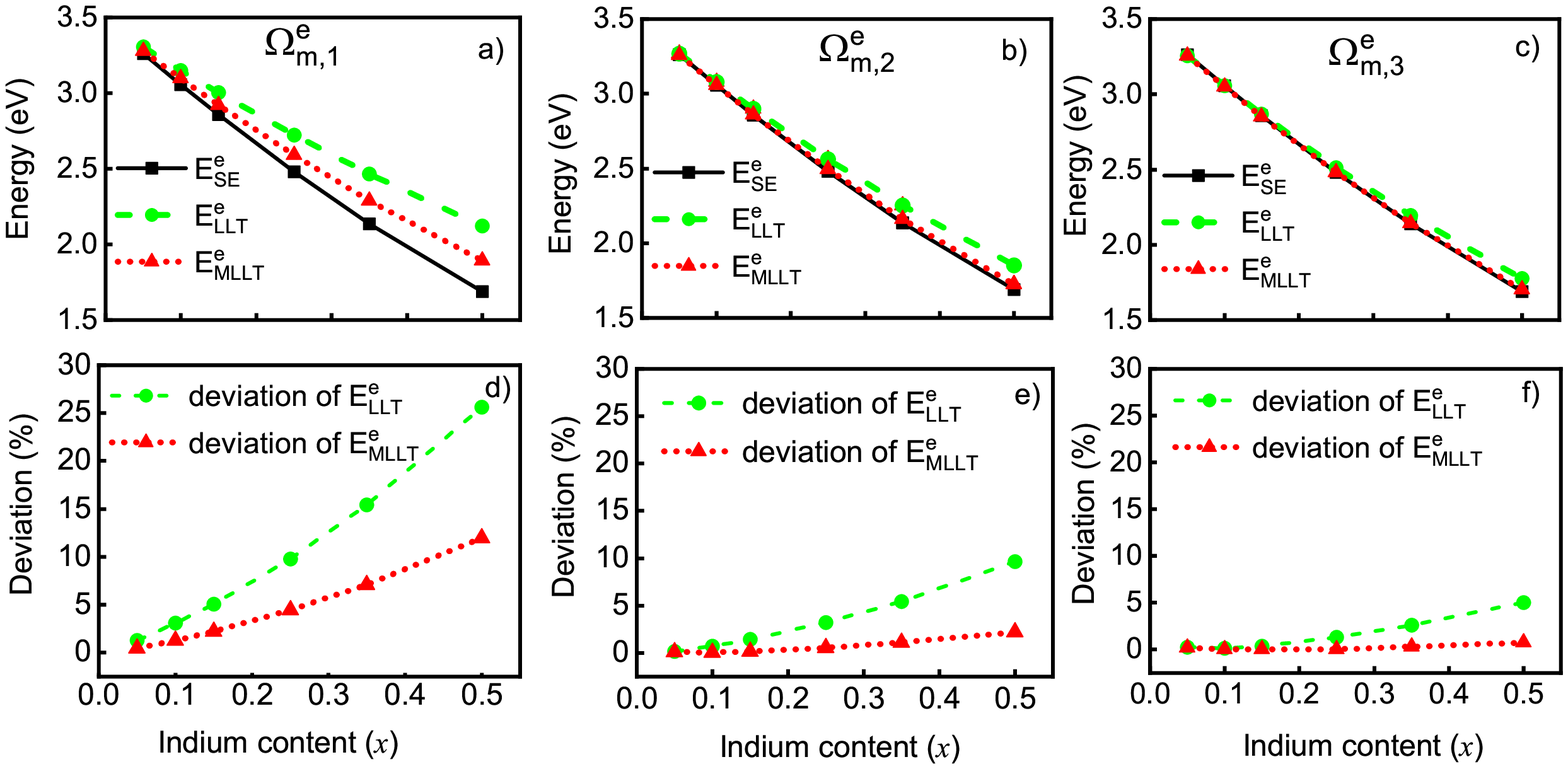}
\caption{Upper row: Electron ground state energies for $c$-plane
In$_x$Ga$_{1-x}$N/GaN QWs as a function of the In content $x$. The
results are shown for the three different subregions
$\Omega^e_{m,i}$ discussed in the text. The electron ground state
energy is computed by solving the SE (black squares), LLT (green
circles) and MLLT (red triangles). Lower row: Deviation of
E$^e_\textnormal{LLT}$ (green circles) and E$^e_\textnormal{{MLLT}}$
(red triangles) with respect to the solution of the SE for the
different subregions $\Omega^e_{m,i}$.}
\label{fig:Electron_GS_Energies}
\end{figure*}

Several aspects of the following analysis are to be noted. Firstly,
the solution of the SE represents the reference/benchmark for the
results of LLT and MLLT. Secondly, since we are using a single-band
effective mass approximation in the framework of a finite difference
method, we treat electrons and holes separately. In doing so,
especially for LLT and MLLT, care must be taken when defining the
zero of energy. As discussed in Sec.~\ref{sec:Theory}, the expansion
coefficients $\alpha_n$ for constructing $u$ from the eigenstates of
the system are inversely proportional to the corresponding state
energies. Ideally, the zero of energy should be chosen close to the
``true'' ground state energy of the system in a given subregion
$\Omega_m$. By doing so, the expansion coefficient $\alpha_1$ is
then large as compared to higher order terms that have lesser
contributions. As a consequence $u$ is then a very good
approximation of the ground state wave function, resulting also in a
good estimate of the corresponding energy. Here, we always choose
the zero of energy as the minimum energy in the band edge profile of
the confining potential for electrons and holes. An illustration of
this situation is displayed in Fig.~\ref{fig:Band_edge}.

Finally, following Eq.~(\ref{eq:Energy_LLT}), when calculating the
ground state energy from $u$, the subspace region $\Omega_m$ over
which $u$ is integrated has to be chosen. To illustrate the impact
of $\Omega_m$ on the results, three different subregions have been
considered for electrons and holes. For electrons these are labeled
as $\Omega^e_{m,1}$, $\Omega^e_{m,2}$ and $\Omega^e_{m,3}$. The
first electron subregion, $\Omega^{e}_{m,1}$, corresponds to the
entire simulation cell (-10 nm to 13.5 nm). The subregion
$\Omega^e_{m,2}$ considers slightly more than the QW region, i.e.
\mbox{-1.5 nm $\leq z\leq$ 5 nm}. For the last subregion,
$\Omega^e_{m,3}$, we just consider it to be \mbox{0 nm $\leq z\leq$
4.5 nm}, meaning that this region starts at the well-barrier
interface at $z=0$ and extends 1 nm into the barrier region above
the upper QW interface at $z=3.5$ nm. This asymmetry in
$\Omega^e_{m,3}$ accounts for the tilt in the band edges and that
therefore the electron wave function is expected to leak further
into the barrier region on the $+z$-side of the well. For holes, the
subregions one $\Omega^h_{m,1}$ and two $\Omega^h_{m,2}$ are
identical to the first two electron cases. Only subregion
$\Omega^h_{m,3}$ is different from $\Omega^e_{m,3}$. For
$\Omega^h_{m,3}$ we have chosen \mbox{-1 nm $\leq z\leq$ 3.5 nm},
which reflects that the tilt in the band edges shifts electron and
hole states in opposite directions.

The aim of using these three different subregions is twofold. First,
as mentioned already above, it will allow us to analyze the impact
of the subregion choice on the ground state energies obtained from
LLT and MLLT in comparison to the result obtained from solving the
SE. Secondly, this analysis also enables us to study if the choice
of $\Omega_m$ affects differently the results obtained from LLT and
MLLT. This insight is for instance of interest when treating
(random) potential fluctuations, where partitioning the system may
be difficult. Thus, a method where results are less dependent on the
$\Omega_m$ choice is in general preferred.

\subsection{Electron ground state energies, wave functions and effective potential}

In a first step we focus on the results for the electron ground
state energy as a function of the In content $x$ for the above
discussed $c$-plane In$_{x}$Ga$_{1-x}$N/GaN QWs. The data are
presented in Fig.~\ref{fig:Electron_GS_Energies}, upper row, for the
three different integration regions $\Omega^e_{m,i}$. The results
obtained by solving the SE, $E^e_\text{SE}$, are given by the black
squares. The green circles show the results from LLT,
$E^e_\text{LLT}$, while the red triangles denote MLLT data. Before
looking at the results in detail, one can already infer from
Fig.~\ref{fig:Electron_GS_Energies} that when using subregion
$\Omega^e_{m,3}$, cf. Fig.~\ref{fig:Electron_GS_Energies} c), a very
good agreement between LLT, MLLT and SE is achieved. Clearly larger
deviations are observed for LLT and MLLT with respect to the SE
result when using the full simulation cell, $\Omega^e_{m,1}$, cf.
Fig.~\ref{fig:Electron_GS_Energies} a).

The lower row of Fig.~\ref{fig:Electron_GS_Energies} displays the
deviations (in \%) between LLT (MLLT) and the solution of the SE as
a function of the In content $x$ for the three different subregions
$\Omega^e_{m,i}$. The green circles show the data for the comparison
between the SE and LLT, while the red triangles do so for the
comparison between SE and MLLT. Starting with $\Omega^e_{m,1}$,
Fig.~\ref{fig:Electron_GS_Energies} d), we observe that for both LLT
and MLLT the deviations increase with increasing In content $x$ in
the well. However, LLT shows noticeably larger deviations when
compared to MLLT for In contents $x$ exceeding values of 15\%
($x=0.15$). Turning to $\Omega^e_{m,2}$,
Fig.~\ref{fig:Electron_GS_Energies} e), deviations are in general
strongly reduced. Nevertheless, still a noticeable difference
between LLT and MLLT is observed. More specifically, while LLT
produces errors of above 5\%, the MLLT results are a good
approximation of the true ground state energy, independent of the In
content $x$ (errors below 2\% are found). When restricting the
integration region further ($\Omega^e_{m,3}$),
Fig.~\ref{fig:Electron_GS_Energies} f), deviations in LLT are
further reduced and only in the very high In content regime
($x>0.4$) more pronounced deviations are observed. The error in the
MLLT result is below 1\% over the range of In composition
considered. This analysis shows that the MLLT produces a better
approximation of the electron ground state energy, independent of
the In content $x$ and chosen subregion $\Omega^e_{m,i}$. This makes
it therefore very attractive for calculations of the fundamental
state in subregion of an energy landscape which shows strong
fluctuations so that the system cannot be easily partitioned into
different subregions.

\begin{figure}[t!]
\includegraphics[width=8.6cm, clip=true]{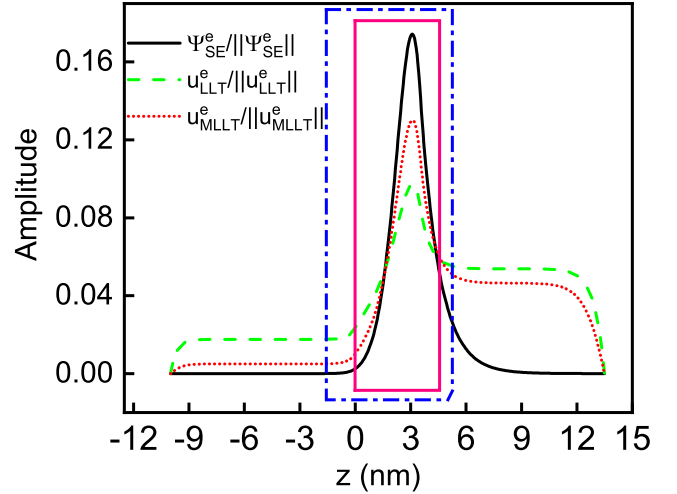}
\caption{(Color online) Comparison of the (normalized) electron
ground state wave functions of a $c$-plane InGaN/GaN QW with 25$\%$
In and a width of 3.5 nm. The wave functions are obtained by solving
the SE (solid black line), LLT (dashed green line) and MLLT (dotted
red line). The blue dashed-dotted box and the solid magenta box
indicate the subregions $\Omega^e_{m,2}$ $\Omega^e_{m,3}$, discussed
in the text.} \label{fig:Electron_WF}
\end{figure}

This begs the question why the energy obtained from MLLT is more
robust against changes in $\Omega_m$. To address this point,
Fig.~\ref{fig:Electron_WF} shows the (normalized) electron ground
state wave function calculated from the SE (black solid line) and
the (normalized) $u$ functions obtained from LLT (green dashed line)
and MLLT (red dotted line) using the full simulation box
$\Omega^e_{m,1}$. The results are displayed for the $c$-plane
In$_{0.25}$Ga$_{0.75}$N/GaN QW. The electron ground state wave
function $\psi^e_\text{SE}$ shows the expected behavior of having
the highest value in the QW region, with the wave function amplitude
then decaying rapidly in the GaN barrier region. Turning to the
result from LLT (green dashed line) first, we find that $u$ has a
maximum in the well, however, it has also a constant finite value in
the GaN barrier region, especially for $z>5$ nm. Thus when changing
the integration region $\Omega^e_m$, a strong impact on the obtained
energy $E_\text{LLT}$ could be expected since contributions from $u$
in the barrier are removed when reducing the subregion
$\Omega^e_{m}$. This is exactly the situation we observe in
Fig.~\ref{fig:Electron_GS_Energies}. More specifically changing the
subregion from $\Omega^e_{m,1}$ (full system) to $\Omega^e_{m,3}$
(mainly QW region), the error in $E_\text{LLT}$ when compared to
$E_\text{SE}$, reduces from 9.8\% to 1.3\% for the
In$_{0.25}$Ga$_{0.75}$N/GaN QW.

\begin{figure*}[t!]
\centering
\includegraphics[width=\textwidth, clip=true]{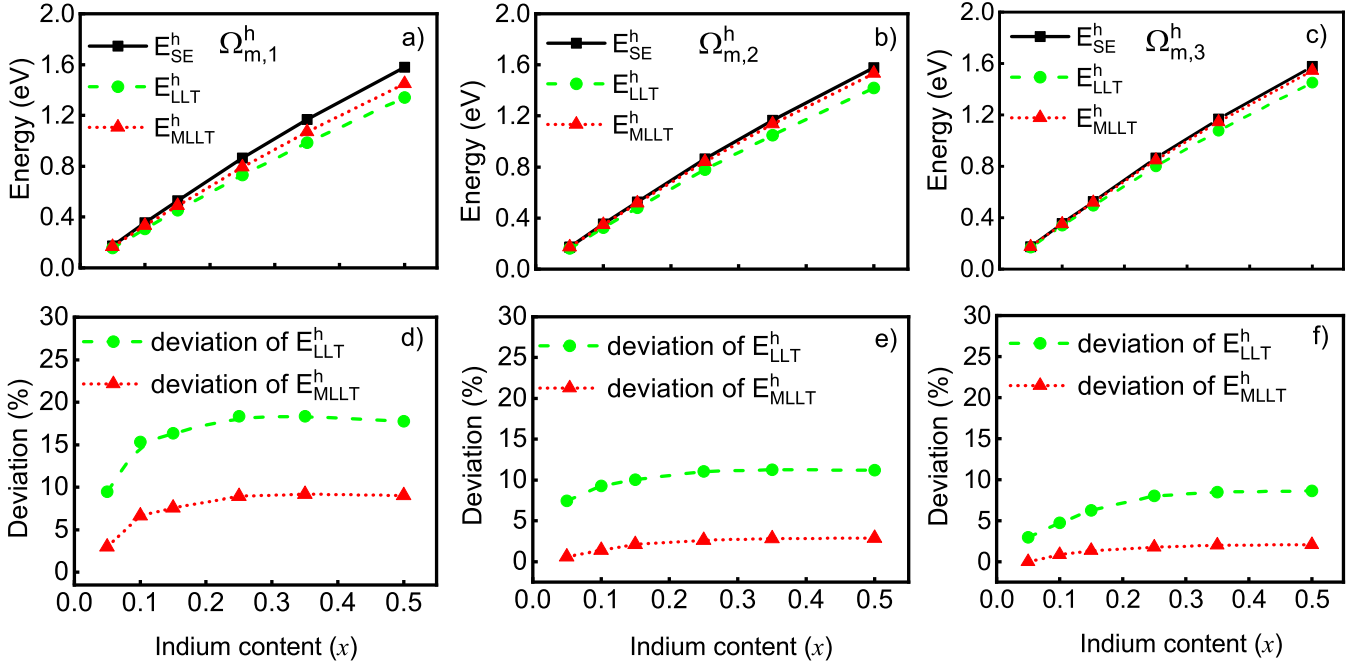}
\caption{Upper row: Hole ground state energies for $c$-plane
In$_x$Ga$_{1-x}$N/GaN QWs as a function of the In content $x$. The
results are shown for the three different subregions
$\Omega^h_{m,i}$ discussed in the text. The hole ground state energy
is computed by solving the SE (black squares), LLT (green circles)
and MLLT (red triangles). Lower row: Deviation of
E$^h_\textnormal{LLT}$ (green circles) and E$^h_\textnormal{{MLLT}}$
(red triangles) with respect to the solution of the SE for the
different subregions $\Omega^h_{m,i}$.} \label{fig:Hole_GS_Energies}
\end{figure*}

The situation is different in the MLLT approach. Here,
$u^e_\text{MLLT}$, at least for $z<0$ nm, gives a better
approximation of $\psi^e_\text{SE}$, with the magnitude of
$u^e_\text{MLLT}$ being very small, similar to $\psi^e_\text{SE}$
but in contrast to $u^e_\text{LLT}$. However, for $z>5$ nm the
magnitude of $u^e_\text{MLLT}$ is comparable to that of
$u^e_\text{LLT}$ and therefore much larger than $\psi^e_\text{SE}$
in this region. Thus, given that the magnitude of $u^e_\text{MLLT}$
is small in the region $z<0$ nm and shows to be a good approximation
of $\psi^e_\text{SE}$ in the QW region, the analysis confirms the
observation that $E^e_\text{MLLT}$ is less dependent on $\Omega^e_m$
than $E^e_\text{LLT}$.

Finally, we discuss here the effective confining potential for
electrons obtained both within LLT, $W^e_\text{LLT}$, and MLLT,
$W^e_\text{MLLT}$. In case of LLT it is obtained via
\mbox{$W^e_\text{LLT}=(u^e_\text{LLT})^{-1}$}, and given in
Fig.~\ref{fig:Band_edge} a) by the green dashed line for an
InGaN/GaN QW with 25\% In. The potential reveals a softening at the
QW barrier interface, which, as discussed above, is an important
feature for quantum corrections in drift-diffusion calculations
using $W^e_\text{LLT}$ for the energy landscape. The here observed
potential profile is consistent with the results reported
previously.~\cite{FiPi2017} To obtain the effective confining
potential from MLLT that reflects the behavior of $W^e_\text{LLT}$,
we find that $W^e_\text{MLLT}=(\sqrt{u^e_\text{MLLT}})^{-1}$ works
here best, while
$\widetilde{W}^e_\text{MLLT}=({u^e_\text{MLLT}}\cdot E_1)^{-1}$
results in a very different effective potential from
$W^e_\text{LLT}$ (not shown). Figure~\ref{fig:Band_edge} a) confirms
that $W^e_\text{MLLT}$ (red dotted line) is in very good agreement
with $W^e_\text{LLT}$ (green dashed line). We note that this holds
over the full In content $x$ range studied here.

\subsection{Hole ground state energies, wave functions and effective potential}

Having discussed the results for the electron ground state energies,
wave functions and the effective confining potential, we now turn
and present the results for holes, again as a function of the In
content $x$ for the above discussed $c$-plane
In$_{x}$Ga$_{1-x}$N/GaN QWs. The upper row of
Fig.~\ref{fig:Hole_GS_Energies} presents the comparison between the
energies obtained from SE, ($E^{h}_\text{SE}$, black squares), LLT,
($E^{h}_\text{LLT}$, green circles) and MLLT, ($E^e_\text{MLLT}$,
red triangles). The results are shown for the three different
subregions $\Omega^h_{m,i}$ over which $u$ is integrated to obtain
the corresponding energy. The lower row of
Fig.~\ref{fig:Hole_GS_Energies} depicts for $\Omega^h_{m,i}$ the
deviation (in \%) of LLT (green circles) and MLLT (red triangles)
from the SE solution. Looking at Fig.~\ref{fig:Hole_GS_Energies} a)
first, one can clearly see that when integrating over the full
simulation region ($\Omega^h_{m,1}$), both $E^{h}_\text{LLT}$ and
$E^{h}_\text{MLLT}$ deviate from $E^{h}_\text{SE}$ with increasing
In content $x$. However, deviations are larger for LLT than for
MLLT. A similar behavior was also observed for the electron ground
state energies when the full simulation box $\Omega^e_{m,1}$ is
considered, cf. Fig.~\ref{fig:Electron_GS_Energies}. But, trends for
electrons and holes are quite different. For electrons, the
deviation in the ground state energies with respect to the SE
solution increase with increasing In content $x$, cf.
Fig.~\ref{fig:Electron_GS_Energies} d). For the holes, deviations in
the ground state energy also start to increase with increasing In
content $x$ but deviations saturate at around 18\% and 8\% for LLT
and MLLT, respectively, when the In content exceeds 15\% ($x>0.15$).
This analysis shows, similar to the results for the electrons, that
when using the full simulation cell ($\Omega^h_{m,1}$), MLLT
provides a better description of $E^h_\text{SE}$ when comparing
errors with LLT. When adjusting/reducing the subregion $\Omega^h_m$,
cf. Figs.~\ref{fig:Hole_GS_Energies} b) and c), to calculate
$E^h_\text{LLT}$ and $E^h_\text{MLLT}$, the agreement with
$E^h_\text{SE}$ clearly improves. This is in particular true for
$E^h_\text{MLLT}$, as Fig.~\ref{fig:Hole_GS_Energies} e) and f)
show; deviations from $E^h_\text{SE}$ close to 3\% or less are
observed over the full In content range $x$. More specifically, for
a well with 25\% In, the error in $E^h_\text{MLLT}$ reduces from 8\%
to 1.8\% (see Figs.~\ref{fig:Hole_GS_Energies} d) and f)) when
changing from $\Omega^h_{m,1}$ to $\Omega^h_{m,3}$. Looking at
$E^h_\text{LLT}$ for the same situation, we observe that the
deviations are reduced from 18.4\% ($\Omega^h_{m,1}$) to 8\%
($\Omega^h_{m,3}$). However, the values are still noticeably higher
when compared to $E_\text{MLLT}$. This also shows that MLLT results
are robust against changes in the In content $x$, while LLT exhibits
larger deviations from the SE data, especially for higher In
contents.

\begin{figure}[t!]
\includegraphics[width=8.6cm, clip=true]{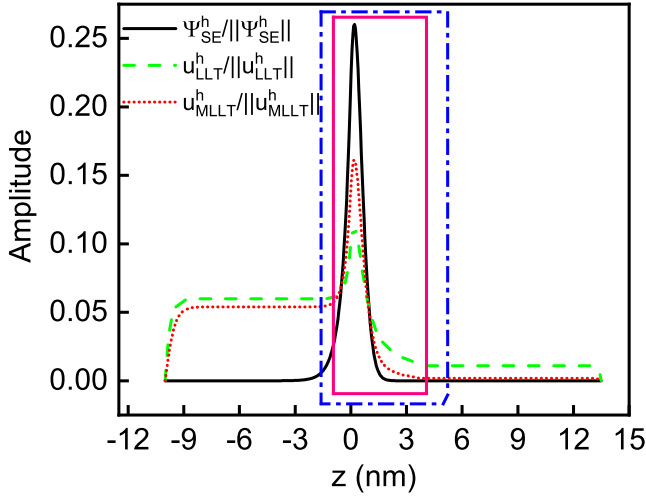}
\caption{(Color online) Comparison of the (normalized) ground state
hole wave functions for a $c$-plane InGaN/GaN QW with 25$\%$ In
content and a width of 3.5 nm. The wave functions are obtained by
solving the SE (solid black line), LLT (dashed green line) and MLLT
(dotted red line), using the simulation box $\Omega^h_{m,1}$. The
blue dashed-dotted box and the solid magenta box represents the two
integration regions from -1.0 nm to 5nm ($\Omega^h_{m,2}$) and -1.0
to 3.5 nm ($\Omega^h_{m,3}$).} \label{fig:hole_WF}
\end{figure}

Following our investigations on the electron ground state energies
and wave functions, we study here also the hole ground state wave
functions. Again we use as a test system the $c$-plane
In$_{0.25}$Ga$_{0.75}$N/GaN QW. The wave functions calculated from
SE (black solid line), LLT (green dashed line) and MLLT (red dotted
line) are shown in Fig.~\ref{fig:hole_WF}. Before looking at the
fine details, independent of the model used, the wave functions are
localized inside the well and ``decay'' in the GaN barrier region.
However, how the wave functions decay in the barrier region strongly
depends on the model. While the hole ground state wave function
$\psi^h_\text{SE}$ rapidly decays in the barrier material, this
situation is only true for $u^h_\text{LLT}$ and $u^h_\text{MLLT}$
along the $+z$-direction. Even though in the $+z$-direction
$u^h_\text{LLT}$ and $u^h_\text{MLLT}$ are similar, there are also
differences. While $u^h_\text{MLLT}$ is very close to 0 in the
barrier region, $u^h_\text{LLT}$ is small, but has a noticeable
finite constant value in the GaN barrier for $z>0$ nm. This effect
is amplified for $z<0$ nm. Again and similar to the electrons, we
attribute differences in the hole ground state energies to
differences in $u$ calculated from MLLT and LLT, given that
deviations in the ground state energy increase as integration region
$\Omega^h_{m,i}$ is increased.

In the last step we turn attention to the effective potential for
holes, $W^h$, calculated within LLT and MLLT. The results are shown
in Fig.~\ref{fig:Band_edge} b). The confining potential from LLT,
\mbox{$W^h_\text{LLT}=(u^h_\text{LLT})^{-1}$}, is given by the green
dashed line and shows again a softening of the potential near the
well barrier interface. Also for holes we have tested calculating
the effective potential from MLLT via
$W^h_\text{MLLT}=\left(\sqrt{u^h_\text{MLLT}}\right)^{-1}$ and
\mbox{$\widetilde{W}^h_\text{MLLT}=\left(E^h_1\cdot{u^h_\text{MLLT}}\right)^{-1}$}.
The conclusion that is drawn from this is similar to that for
electrons, meaning that $\widetilde{W}^h_\text{MLLT}$ gives a
potential profile very different from $W^h_\text{LLT}$ (not shown),
while $W^h_\text{MLLT}$ is in good agreement with $W^h_\text{LLT}$.
This is confirmed by Fig.~\ref{fig:Band_edge} b), showing that the
confining potential obtained from MLLT (red dotted line) captures
the same effects as $W^h_\text{LLT}$ (green dashed line). Again this
result holds over the full In content range investigated here.

\section{Conclusions}
\label{sec:Conclusion}

In this work we have proposed, motivated and analyzed, a modified
localization landscape theory (MLLT). In the MLLT approach we solve
$\hat{H}^2u=1$ instead of $\hat{H}u=1$, as in the LLT. We
demonstrate the improvements resulting from using $\hat{H}^2u=1$ in
predicting ground state energies for a particle-in-a-box (infinite
square well) potential. Since this problem can be solved fully
analytically in LLT and MLLT, the solution confirms that $u$
obtained from MLLT will in general give a better approximation of
the true ground state wave function when compared to the result from
LLT. We have also shown that this can be traced back to the energy
dependence of the expansion coefficients of $u$ in terms of the
particle-in-a-box eigenstates. Given that $u$ obtained from MLLT
provides a very good description of the ground state wave function,
it also provides an improved estimate of the ground state energy and
therefore the error in this quantity is reduced in comparison to the
LLT obtained value. Here, we also have provided insight into the
calculation of the effective confining potential $W$ within MLLT.
While this is straightforward in the case of a particle-in-a-box
problem with infinitely high barriers, we highlight that care must
be taken when extracting the effective confining $W$ from MLLT in
general. We have discussed two strategies to obtain $W$ from MLLT
that lead to results similar to those obtained from LLT, which is
important when applying MLLT for instance in drift-diffusion
transport calculations to account for quantum corrections.

The particle-in-a-box problem provided the ideal testbed to study
the basic properties of the LLT and MLLT. However, further analysis
is required to consider more realistic potentials. LLT has recently
been used to evaluate the electronic structure of III-N
heterostructures, where the confining potential is triangular shaped
with barriers of finite height. Motivated by this, we have studied
and compared ground state energies from the Schr\"{odinger equation}
(SE), LLT and MLLT for $c$-plane In$_{x}$Ga$_{x}$N/GaN QWs as a
function of the In content $x$. Special attention was  paid to the
impact of the choice of the integration region of $u$ when
evaluating the ground state energies. Our calculations reveal that
for both electron and hole ground states, MLLT always gives a better
description of the ``true'' ground state energy when compared to the
LLT result. We also find that the subregion over which $u$ is being
integrated to obtain this energy is less important for MLLT than it
is for LLT. Over the composition range from 5\% to 50\% In in the
well and when integrating over a region close to the QW, errors in
the ground state energy from MLLT never exceeds 4\%. While similar
numbers are obtained for LLT in the lower In content range ($<$15\%
In) and when choosing appropriate subregions, especially for holes
at higher In contents ($>$25\% In), errors in the range of 5\% to
10\% are observed. Looking at the calculated effective potential $W$
for electrons and holes, and independent of the In content $x$, we
find that using $W_\text{MLLT}=(\sqrt{u_\text{MLLT}})^{-1}$ gives in
general results that match closely the effective potential
$W_\text{LLT}$ obtained from LLT. Since $W$ plays an important role
in quantum corrected drift-diffusion simulations, it is useful to
see that MLLT produces an energy landscape similar to LLT, so that
it can be used in such simulations.

Taking all this together the proposed MLLT keeps all the benefits of
the LLT, such that only a system of linear equations has to be
solved instead of a large eigenvalue problem to obtain ground state
energies. At the same time the MLLT provides the following aspects:
(i) ``faster convergence'' of the calculated ground state energies
with integration region, (ii) a more ``robust'' behavior of the
method against changes in the integration region, (iii) better
agreement with results from SE, especially for higher In contents
and (iv) an effective confining potential comparable to that of LLT.
All these features make the MLLT method attractive for calculations
of localized states in highly disordered systems, where for instance
partitioning the systems into different subregions is not trivial.

\begin{acknowledgements}
The authors acknowledge financial support from Science Foundation
Ireland  under Grant Nos. 17/CDA/4789, 15/IA/3082 and 12/RC/2276 P2.
\end{acknowledgements}

\appendix*
\section{Infinite Triangular Well}
\label{appendixI}

Having discussed in the main text the fully analytic solution of the
particle-in-a-box problem with infinitely high barriers, we study
here another problem that can be investigated fully analytically,
which is a triangular well with infinitely high barrier at $z = 0$;
the potential increases from $0$ at $z=0$ with a slope $F$ in the
+$z$-direction. The aim of this study is twofold: it will illustrate
(i) that in contrast to the particle-in-a-box problem, discussed in
Sec.~\ref{sec:PIAB}, the expansion coefficients of $u$ can depend on
multiple energies $E_n$ and (ii) that there are potentials where LLT
and MLLT could fail to give a good approximation of (ground state)
energies or even diverge. Regarding (i), this finding is important
for calculating the effective confining potential, showing that it
might not always be guaranteed that calculating
$\widetilde{W}_\text{MLLT}=(E_1\cdot u_\text{MLLT})^{-1}$ will give
a good approximation of $W_\text{LLT}$ obtained from the
``standard'' LLT approach.

For the infinite triangular potential, the SE reads:
\begin{eqnarray}
 -\frac{\hbar^2}{2m}\frac{d^2}{dz^{2}}\psi_{n}(z) + Fz\psi_{n}(z) =
 E_{n}\psi_{n}(z)\nonumber\\
\Leftrightarrow \frac{d^2}{dz^{2}}\psi_{n}(z) -
\frac{2mF}{\hbar^2}\left(z - \frac{E_n}{F}\right)\psi_{n}(z)= 0 \,\,
. \label{31}
\end{eqnarray}
Setting $a = (2mF/\hbar^2 )^{1/3}$ and using $\gamma = a(z-E_n/F)$,
one is left with
\begin{equation}
a^2\left[\frac{d^2f(\gamma)}{d\gamma^{2}} -\gamma
f(\gamma)\right]=0\,\, . \label{34}
\end{equation}
The general solution to the above differential equation can be
obtained as a linear combination of the Airy functions $A(z)$ and
$B(z)$. These functions are defined as the improper Riemann
integrals~\cite{VaSo2004}
\begin{equation}
A(z) = \frac{1}{\pi}\lim_{c\to\infty}\int_{0}^{c} \cos\left(\frac{t^3}{3}+zt\right)dt\,\, ,\\
\label{35}
\end{equation}
\begin{equation}
B(z)=\frac{1}{\pi}\lim_{h\to\infty}\int_{0}^{h}\left[\exp\left(\frac{t^3}{3}+zt\right)
+ \sin\left(\frac{t^3}{3}+zt\right)\right]dt\,\, . \label{36}
\end{equation}
\\
For $z > 0$, the function $A(z)$ shows exponential decay whereas
$B(z)$ diverges to infinity. Given that the confined wave functions
have to decay as $z \rightarrow \infty$, the function $B(z)$ has to
be discarded. Thus, the solutions of the SE for a triangular-shaped
potential with infinitely high barriers, cf. Eq.~(\ref{31}), are
given by:
\begin{equation}
\psi_n(z) = \alpha_n
A_i\left(\left(\frac{2mF}{\hbar^2}\right)^{1/3}\left(z
-\frac{E_n}{F}\right)\right)\,\, . \label{37}
\end{equation}
The fact that the wave function has to go to zero at the infinitely
high barrier at $z = 0$ can be used to determine the energy
eigenvalues $E_n$. To do so, the $n$th zero of the Airy function is
approximated and the corresponding eigenvalue then reads:
\begin{equation}
E_n \approx \left(\frac{3\pi F
\hbar}{8m^2}\left(n-\frac{1}{4}\right)\right)^{2/3}\,\, . \label{38}
\end{equation}

Equipped with this solution we turn now and discuss the infinite
triangular well firstly in the framework of LLT and then of MLLT. To
find a series expansion for $u$, we first need an expansion for the
constant function 1 in terms of the eigenfunctions, over the
interval [0, $\infty$):~\cite{AbSt1965}
\begin{equation}
\sum_{n=1}^{\infty}b_n\psi_n = 1\,\, . \label{39}
\end{equation}
We recall here that $u =\sum_{n=1}^{\infty}a_n\psi_n$,
Eq.~(\ref{eq:u_def}), so that when using LLT, $\hat{H}u=1$, one is
left with
\begin{equation}
\hat{H}u = \hat{H}\left(\sum_{n=1}^{\infty}a_n\psi_n\right) =
\left(\sum_{n=1}^{\infty}a_nE_n\psi_n\right) = 1 =
\sum_{n=1}^{\infty}b_n\psi_n\,\, . \label{40}
\end{equation}
Due to the orthonormality of the wave functions, we can thus express
the expansion coefficients $a_n$ as
\begin{equation}
a_n = \frac{b_n}{E_n}\,\, . \label{41}
\end{equation}
Thus, within LLT, $u$ can be expressed as:
\begin{equation}
u_\textnormal{LLT}(z) = \frac{b_1}{E_1}\psi_1(z) + \frac{b_2}{E_2}\psi_2(z) +  \frac{b_3}{E_3}\psi_3(z) + \cdot\cdot\cdot
\label{42}
\end{equation}
In contrast to the infinite square-well potential, one can show that
the sum of the $a_n$'s does not converge for the triangular well
considered and thus the energy $E^{(u)}$,Eq.~(\ref{eq:Energy_LLT}),
does not converged. First, we note that the energies $E_n$,
Eq.~(\ref{38}), increase with increasing $n$ as $n^{\frac{2}{3}}$;
numerical analysis we have undertaken indicates that $b_n$ decays
approximately as $n^{-\alpha}$, where $\alpha\approx0.20$; hence
$a_n$ in LLT decays approximately as $n^{-(\alpha+\frac{2}{3})}$,
where $\alpha+\frac{2}{3}<1$, which results in a divergent series
for $u$, Eq.~(\ref{eq:u_def}).



Turning to the MLLT, and following the same procedure as for the
LLT, we find here that the coefficients $a_n$ are given by
\begin{equation}
a_n = \frac{b_n}{E^2_n}\,\, . \label{eq:MLLT_tringular}
\end{equation}
Successive terms in the series Eq.~(\ref{eq:MLLT_tringular}) decay
as $n^{-(\alpha+\frac{4}{3})}$. Therefore, higher order term
contributions are reduced in the MLLT case. Numerical studies
confirm that MLLT converges with increasing system size, in contrast
to the LLT. However, MLLT converges to a ground state energy that is
noticeably larger than the ground state energy obtained from SE. All
this highlights benefits but also potential shortcomings or problems
in both LLT and MLLT for certain potentials.

\bibliographystyle{apsrev}

\end{document}